\documentclass[aip,apl,reprint]{revtex4-1}

\usepackage{graphicx}
\usepackage{dcolumn}
\usepackage{bm}
\usepackage{color}
\usepackage{verbatim} 

\begin{document}

\title{Lateral imaging of the superconducting vortex lattice using Doppler-modulated scanning tunneling microscopy} 

\author{I. Fridman}
\affiliation{Department of Physics, University of Toronto, 60 St. George St., Toronto ON M5S1A7 Canada}

\author{C. Kloc}
\affiliation{School of Materials Science and Engineering, Nanyang Technological University 639798 Singapore}

\author{C. Petrovic}
\affiliation{Condensed Matter Physics and Materials Science Department, Brookhaven National Laboratory, Upton, NY 11973 USA}

\author{J. Y. T. Wei}
\affiliation{Department of Physics, University of Toronto, 60 St. George St., Toronto ON  M5S1A7 Canada}
\affiliation{Canadian Institute for Advanced Research, Toronto, ON, M5G1Z8 Canada}

\begin{abstract}

By spatially mapping the Doppler effect of an in-plane magnetic field on the quasiparticle tunneling spectrum, we have laterally imaged the vortex lattice in superconducting $2H$-NbSe$_2$.  Cryomagnetic scanning tunneling spectroscopy was performed at 300 mK on the $ab$-surface oriented parallel to the field $H$.  Conductance images at zero bias show stripe patterns running along $H$, with the stripe separation varying as $H^{-0.5}$.  Regions of higher zero-bias conductance show lower gap-edge conductance, consistent with spectral redistribution by spatially-modulated superfluid momentum.  Our results are interpreted in terms of the interaction between vortical and screening currents, and demonstrate a general method for probing subsurface vortices.

\end{abstract}


\maketitle

In response to an applied magnetic field, type-II superconductors experience a diamagnetic current that circulates along the sample edge.  Above the lower critical field, field can penetrate into the superconductor via a lattice of vortices, each consisting of a paramagnetic current loop enclosing a flux quantum. \cite{Abrikosov}  The vortex lattice can be imaged using techniques sensitive to variations in the local magnetic field such as Bitter decoration\cite{Essmann67} and Lorentz microscopy,\cite{Matsuda01} or with scanning tunneling microscopy (STM) which probes the local quasiparticle density of states (DOS).  STM imaging of the vortices is possible by virtue of bound states and suppressed superconducting gap in the vortex core.\cite{Caroli64, Hess89}  Because of this reliance on vortex-core states, STM imaging has been largely limited to the cross-sectional geometry, i.e. with the vortices piercing the sample surface.  An earlier STM study versus field direction has shown the density of vortices to decrease as the field is tilted away from the surface normal.\cite{Hess92, Hess94}  For fields parallel to the surface, the vortex cores become buried in the bulk, making them difficult to probe directly.  In this lateral field geometry, vortex lattices have been imaged by Lorentz microscopy, but only in highly 2D superconductors where pancake vortices decorate in-plane flux lines.\cite{Vlasov02,Tamegai02} 

In this letter, we report on lateral imaging of the superconducting vortex lattice using cryomagnetic scanning tunneling spectroscopy.  By mapping the zero-bias tunneling conductance over the $ab$-surface of superconducting $2H$-NbSe$_2$ in an in-plane magnetic field and at 300 mK, we have observed distinct stripe patterns whose orientation and spacing versus the field can be directly attributed to the in-plane flux lattice.  Our observations are interpreted in terms of the interaction between the diamagnetic screening current and the paramagnetic vortical currents, which results in a spatially-modulated Doppler effect on the quasiparticle DOS spectrum.

\begin{figure}[b]
\includegraphics {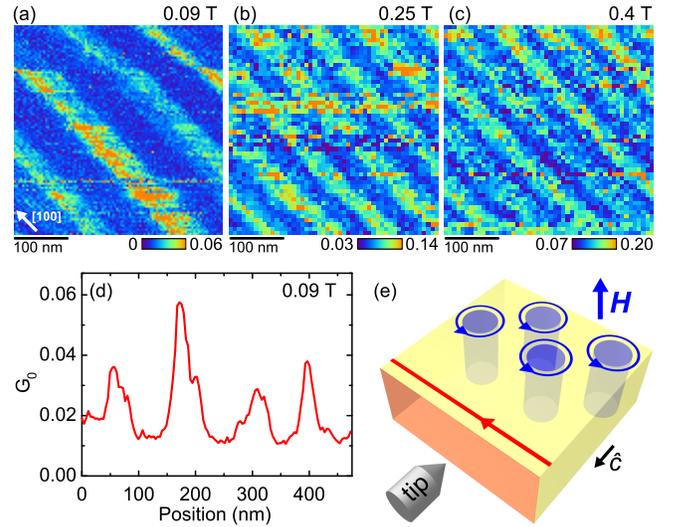}
\caption{\label{fig1} Lateral imaging of the superconducting vortex lattice in $2H$-NbSe$_2$ at various fields and 300 mK. Panels (a) to (c) show 380 x 380 nm$^2$ spatial maps of the normalized zero-bias tunneling conductance $G_0$ for fields of 0.09 T, 0.25 T, and 0.4 T, applied along the [100] direction (white arrow in (a)). Panel (d) shows the average $G_0$ along the direction perpendicular to the stripes in the 0.09 T data.  Panel (e) shows a schematic of our experiment, with the STM measuring the $ab$-surface, across which flows a diamagnetic screening current (red line).  In the heuristic model discussed in the text, paramagnetic currents (blue loops) circulating the subsurface vortices perturb the screening current, thus spatially modulating $G_0$ and producing the observed $G_0$ image contrast.}  \end{figure}

The STM used in our experiment was specially designed for the magnetic field to be applied parallel to the sample surface, as shown in Fig. \ref{fig1}(e). The STM is mounted inside a $^3$He cryostat which is inserted into a superconducting solenoid.  The Pt-Ir tips used were field-emitted $in$ $situ$ to ensure stable vacuum tunneling, and RF-filters were used throughout the wiring to maximize the spectral resolution. The $dI/dV$ conductance spectra were acquired by lock-in amplification with a 20 $\mu$V excitation at 505 Hz, and the typical high-bias junction impedance was 10 M$\Omega$.

Single crystals of $2H$-NbSe$_2$ were grown by an iodine vapor transport technique.\cite{Oglesby94}  The crystals had critical temperatures of $\approx$ 7.2 K, and upper critical fields of $\approx$ 5 T and 15 T respectively for field perpendicular and parallel to the $ab$-plane.  The crystals measured were $\sim$ 5 x 5 x 0.5 mm$^3$ in size, with the wide faces normal to the $c$-axis.  The crystals were oriented by X-ray diffraction, cleaved just before being loaded into the STM and cooled to 300 mK in zero field.  STM topography revealed atomically smooth surfaces, with hexagonally arranged Se atoms modulated by triangular charge density waves.  The magnetic field was applied along the [100] direction with $\sim$ 2$^\circ$ precision.  

Figure \ref{fig1} shows spatial maps of the zero-bias conductance, $G_0$. The $G_0$ data was normalized relative to the above-gap conductance at 4 mV.  Panels (a) to (c) show plots of the data for 0.09 T, 0.25 T and 0.4 T respectively.  Regions of high $G_0$, hereafter referred to as stripes, can be seen in each plot, running parallel to the in-plane field and spaced at regular intervals.  Crystals from different growth batches showed the same stripe patterns, thus attesting to their general reproducibility.  The average half-width of the stripes is 35$\pm$15 nm over the field range measured.  This is comparable to the expected size of a vortex core 2$\xi_{ab}$ $\sim$ 20 nm, where $\xi_{ab}$ is the zero-temperature superconducting coherence length in the $ab$-plane for $2H$-NbSe$_2$. \cite{Sonier05}  The brightness of the stripes alternates between adjacent stripes, an effect more clearly seen in panel (d), which shows a profile plot of the average $G_0$ along the direction perpendicular to the stripes in the 0.09 T data.  

\begin{figure}[t]
\includegraphics {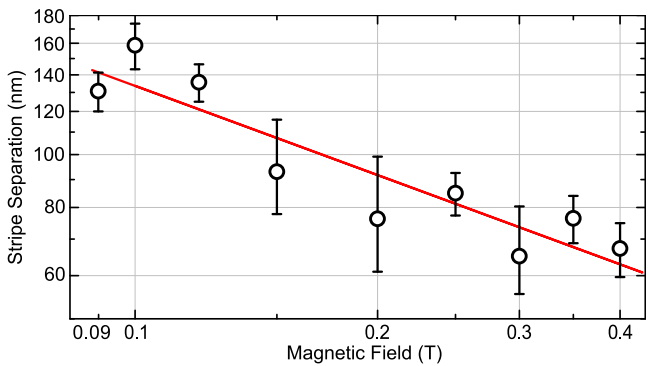}
\caption{\label{fig2} Separation between the centers of the $G_0$ stripes, for fields between 0.09 T and 0.4 T, plotted on a log-log scale.  The data (circles) are fitted to $\propto H^{-0.54\pm0.09}$ (red line), in good agreement with the expected $H^{-0.5}$ dependence of the vortex lattice parameter on field.}
\end{figure}

To interpret the stripe patterns as a manifestation of the subsurface flux vortices, we consider quantitatively how these patterns vary with the applied field $H$.  Figure \ref{fig2} shows a log-log plot of the separation between stripe centers over the field range measured in this experiment.  The stripe separation decreases with increasing field and is fitted to $\propto H^{-0.54\pm0.09}$, in good agreement with the expected $H^{-0.5}$ dependence of the vortex lattice parameter on field.  It is important to note that the stripe separation we observed is 0.86$\pm$0.10 times the Abrikosov lattice parameter.  This observation can be qualitatively explained by identifying every other stripe with the row of flux lines closest to the surface, and the in-between stripes as coming from the next closest row; this picture would be consistent with the alternating brightness between adjacent stripes noted above.  A more quantitative explanation of the stripe separation observed would need to consider vortex lattice distortions due to the superconducting anisotropy of $2H$-NbSe$_2$ over the field range we measured. \cite{Kogan88}

\begin{figure}
\includegraphics {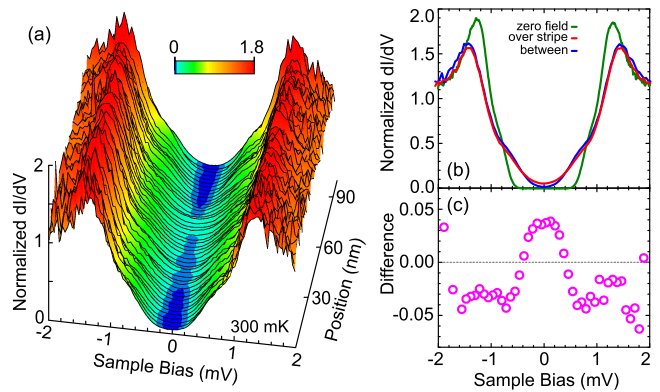}
\caption{\label{fig3} (a) Series of normalized $dI/dV$ spectra measured as the tip is scanned perpendicularly across a stripe at 0.1 T and 300 mK.  The midgap states rise over a $\approx$ 35 nm wide region, thus accounting for the stripe patterns seen in the $G_0$ images.  (b) Spatially-averaged $dI/dV$ spectra over the center of a stripe (red curve) and just between two stripes (blue curve).  The zero-field spectrum (green curve) is included for comparison, to show the overall effect of the applied field.  Over the stripe, the zero-bias $dI/dV$ is higher while the coherence-peak $dI/dV$ is lower, an effect clearly seen in the difference of the two curves, as shown in (c).}
\end{figure}

To visualize spectroscopically how the stripe patterns emerge, we analyze detailed variations in the $dI/dV$ spectrum across a typical $G_0$ image.  Figure \ref{fig3}(a) shows a series of $dI/dV$ spectra measured while the tip is scanned perpendicularly across a stripe region at 0.1 T.  It is clear that the midgap states rise appreciably over a $\approx$ 35 nm wide region, thus accounting for the stripe patterns seen in the $G_0$ image described above.  Panel (b) shows the spatially-averaged spectra over the center of a stripe and just between two stripes.  The zero-field spectrum is included for comparison, to show the overall effect of the applied field.  Comparing the two curves at 0.1 T, the zero-bias $dI/dV$ is higher while the $dI/dV$ near the coherence peak at $\approx$ $\pm$1.4 mV is lower, over a stripe than between stripes.  This effect can be seen more clearly in the difference of the two spectra, as shown in panel (c).  Here we emphasize the subtlety of this midgap-states effect, i.e. the difference in $G_0$ is $\approx$ 5 nS, requiring high measurement sensitivity to detect.

To understand the physical origin of the spectral evolution versus field, we consider the recent theoretical model of Zhang $\emph{et al.}$ (Ref. \onlinecite{Zhang04}), which calculates the quasiparticle DOS spectrum for a current-carrying superconductor.  Essentially, the supercurrent adds a Doppler term proportional to $\bm{v}_F \cdot \bm{q}_s$ to the quasiparticle energy dispersion $E_k$, where $\bm{v}_F$ is the Fermi velocity and $\bm{q}_s$ is the superfluid momentum.  This Doppler shift in $k$-space causes a redistribution of the energy-gap size along the Fermi surface (FS), thus modifying the quasiparticle DOS spectrum.\cite{Fulde69}  With increasing $\bm{q}_s$, the coherence peaks are suppressed in height and shifted to higher energies.\cite{Zhang04, Kohen06}  When the Doppler term becomes sufficiently large relative to the gap amplitude, parts of the FS become gapless, causing the zero-energy DOS to increase with the field.  All of these spectral behaviors, namely suppression of the coherence peak and enhancement of the midgap states, are observed in our tunneling measurements, thus directly implicating the Doppler effect in the appearance of finite $G_0$ with in-plane field.

Finally, we consider the interaction between the vortical and screening currents, in order to explain why the Doppler redistribution of the quasiparticle DOS spectrum shows spatial modulation that is correlated with the vortex lattice.  These two currents flow in opposite directions near the sample surface, as shown in Fig. \ref{fig1}(e), and the counter-flowing current lines would be denser over certain regions in order to avoid crossing each other, thereby locally enhancing the supercurrent density and thus the superfluid momentum.  A spatially-modulated superfluid momentum implies a spatially-modulated Doppler effect, which could then produce the stripe patterns seen in our in-field STM measurements.  In this heuristic model, the stripes in the $G_0$ image corresponding to high-Doppler regions are over vortices, while the spaces between stripes corresponding to low-Doppler regions are between vortices.  Although this model is physically plausible, it should be noted that there are alternative models indicating the opposite scenario, i.e. larger current density and thus stronger Doppler effect between vortices than over vortices.\cite{Brandt81,Hernandez08}  Applicabilities of these models depend on quantitative details of the model assumptions, in particular the field strength relative to the thermodynamic critical field and the coherence length relative to the London penetration depth.  Further theoretical work is needed to more rigorously justify our model and quantitatively interpret our data.  Here it is also worth noting that the subgap features at $\approx$ $\pm$0.7 mV seen in our spectra could be due to the multiband nature of $2H$-NbSe$_2$, as manifestations of a smaller superconducting energy gap which is also Doppler-redistributed by the in-plane field. \cite{Boaknin03,Huang07,Rodrigo04,Guillamon08} Elucidation of this scenario would require a multiband generalization of the Zhang $\emph{et al.}$ model.\cite{Zhang04}, but would not affect the stripe patterns we observed or our interpretation of their physical origin.

In summary, we have used Doppler-modulated scanning tunneling microscopy, performed at 300 mK, to laterally image the in-plane vortex lattice in superconducting $2H$-NbSe$_2$.  This technique can in principle be applied on various other superconductors to probe the length scales and spatial symmetry of subsurface flux vortices.  It may also potentially be used to study other fundamental vortex phenomena, such as single-vortex entry and vortex-surface interactions.

We acknowledge useful discussions with J. R. Clem, C.-R. Hu, V. G. Kogan and A. Paramekanti.  This work was supported by NSERC, CFI-OIT and the Canadian Institute for Advanced Research.  Part of this work was carried out at BNL, which is operated for the U.S. Department of Energy by Brookhaven Science Associates (Grant No. DE-Ac02-98CH10886).


\begin{thebibliography}{21}%
\makeatletter
\providecommand \@ifxundefined [1]{%
 \@ifx{#1\undefined}
}%
\providecommand \@ifnum [1]{%
 \ifnum #1\expandafter \@firstoftwo
 \else \expandafter \@secondoftwo
 \fi
}%
\providecommand \@ifx [1]{%
 \ifx #1\expandafter \@firstoftwo
 \else \expandafter \@secondoftwo
 \fi
}%
\providecommand \natexlab [1]{#1}%
\providecommand \enquote  [1]{``#1''}%
\providecommand \bibnamefont  [1]{#1}%
\providecommand \bibfnamefont [1]{#1}%
\providecommand \citenamefont [1]{#1}%
\providecommand \href@noop [0]{\@secondoftwo}%
\providecommand \href [0]{\begingroup \@sanitize@url \@href}%
\providecommand \@href[1]{\@@startlink{#1}\@@href}%
\providecommand \@@href[1]{\endgroup#1\@@endlink}%
\providecommand \@sanitize@url [0]{\catcode `\\12\catcode `\$12\catcode
  `\&12\catcode `\#12\catcode `\^12\catcode `\_12\catcode `\%12\relax}%
\providecommand \@@startlink[1]{}%
\providecommand \@@endlink[0]{}%
\providecommand \url  [0]{\begingroup\@sanitize@url \@url }%
\providecommand \@url [1]{\endgroup\@href {#1}{\urlprefix }}%
\providecommand \urlprefix  [0]{URL }%
\providecommand \Eprint [0]{\href }%
\providecommand \doibase [0]{http://dx.doi.org/}%
\providecommand \selectlanguage [0]{\@gobble}%
\providecommand \bibinfo  [0]{\@secondoftwo}%
\providecommand \bibfield  [0]{\@secondoftwo}%
\providecommand \translation [1]{[#1]}%
\providecommand \BibitemOpen [0]{}%
\providecommand \bibitemStop [0]{}%
\providecommand \bibitemNoStop [0]{.\EOS\space}%
\providecommand \EOS [0]{\spacefactor3000\relax}%
\providecommand \BibitemShut  [1]{\csname bibitem#1\endcsname}%
\let\auto@bib@innerbib\@empty
\bibitem [{\citenamefont {Abrikosov}(1957)}]{Abrikosov}%
  \BibitemOpen
  \bibfield  {author} {\bibinfo {author} {\bibfnamefont {A.~A.}\ \bibnamefont
  {Abrikosov}},\ }\href@noop {} {\bibfield  {journal} {\bibinfo  {journal}
  {Sov. Phys. JETP}\ }\textbf {\bibinfo {volume} {5}},\ \bibinfo {pages} {1174}
  (\bibinfo {year} {1957})}\BibitemShut {NoStop}%
\bibitem [{\citenamefont {Essmann}\ and\ \citenamefont
  {Träuble}(1967)}]{Essmann67}%
  \BibitemOpen
  \bibfield  {author} {\bibinfo {author} {\bibfnamefont {U.}~\bibnamefont
  {Essmann}}\ and\ \bibinfo {author} {\bibfnamefont {H.}~\bibnamefont
  {Träuble}},\ }\href {\doibase DOI: 10.1016/0375-9601(67)90819-5} {\bibfield
  {journal} {\bibinfo  {journal} {Phys. Lett. A}\ }\textbf {\bibinfo {volume}
  {24}},\ \bibinfo {pages} {526 } (\bibinfo {year} {1967})}\BibitemShut
  {NoStop}%
\bibitem [{\citenamefont {Matsuda}\ \emph {et~al.}(2001)\citenamefont
  {Matsuda}, \citenamefont {Kamimura}, \citenamefont {Kasai}, \citenamefont
  {Harada}, \citenamefont {Yoshida}, \citenamefont {Akashi}, \citenamefont
  {Tonomura}, \citenamefont {Nakayama}, \citenamefont {Shimoyama},
  \citenamefont {Kishio}, \citenamefont {Hanaguri},\ and\ \citenamefont
  {Kitazawa}}]{Matsuda01}%
  \BibitemOpen
  \bibfield  {author} {\bibinfo {author} {\bibfnamefont {T.}~\bibnamefont
  {Matsuda}}, \bibinfo {author} {\bibfnamefont {O.}~\bibnamefont {Kamimura}},
  \bibinfo {author} {\bibfnamefont {H.}~\bibnamefont {Kasai}}, \bibinfo
  {author} {\bibfnamefont {K.}~\bibnamefont {Harada}}, \bibinfo {author}
  {\bibfnamefont {T.}~\bibnamefont {Yoshida}}, \bibinfo {author} {\bibfnamefont
  {T.}~\bibnamefont {Akashi}}, \bibinfo {author} {\bibfnamefont
  {A.}~\bibnamefont {Tonomura}}, \bibinfo {author} {\bibfnamefont
  {Y.}~\bibnamefont {Nakayama}}, \bibinfo {author} {\bibfnamefont
  {J.}~\bibnamefont {Shimoyama}}, \bibinfo {author} {\bibfnamefont
  {K.}~\bibnamefont {Kishio}}, \bibinfo {author} {\bibfnamefont
  {T.}~\bibnamefont {Hanaguri}}, \ and\ \bibinfo {author} {\bibfnamefont
  {K.}~\bibnamefont {Kitazawa}},\ }\href {\doibase 10.1126/science.1065968}
  {\bibfield  {journal} {\bibinfo  {journal} {Science}\ }\textbf {\bibinfo
  {volume} {294}},\ \bibinfo {pages} {2136} (\bibinfo {year}
  {2001})}\BibitemShut {NoStop}%
\bibitem [{\citenamefont {Caroli}, \citenamefont {Gennes},\ and\ \citenamefont
  {Matricon}(1964)}]{Caroli64}%
  \BibitemOpen
  \bibfield  {author} {\bibinfo {author} {\bibfnamefont {C.}~\bibnamefont
  {Caroli}}, \bibinfo {author} {\bibfnamefont {P.~G.~D.}\ \bibnamefont
  {Gennes}}, \ and\ \bibinfo {author} {\bibfnamefont {J.}~\bibnamefont
  {Matricon}},\ }\href {\doibase DOI: 10.1016/0031-9163(64)90375-0} {\bibfield
  {journal} {\bibinfo  {journal} {Phys. Lett.}\ }\textbf {\bibinfo {volume}
  {9}},\ \bibinfo {pages} {307 } (\bibinfo {year} {1964})}\BibitemShut
  {NoStop}%
\bibitem [{\citenamefont {Hess}\ \emph {et~al.}(1989)\citenamefont {Hess},
  \citenamefont {Robinson}, \citenamefont {Dynes}, \citenamefont {Valles},\
  and\ \citenamefont {Waszczak}}]{Hess89}%
  \BibitemOpen
  \bibfield  {author} {\bibinfo {author} {\bibfnamefont {H.~F.}\ \bibnamefont
  {Hess}}, \bibinfo {author} {\bibfnamefont {R.~B.}\ \bibnamefont {Robinson}},
  \bibinfo {author} {\bibfnamefont {R.~C.}\ \bibnamefont {Dynes}}, \bibinfo
  {author} {\bibfnamefont {J.~M.}\ \bibnamefont {Valles}}, \ and\ \bibinfo
  {author} {\bibfnamefont {J.~V.}\ \bibnamefont {Waszczak}},\ }\href {\doibase
  10.1103/PhysRevLett.62.214} {\bibfield  {journal} {\bibinfo  {journal} {Phys.
  Rev. Lett.}\ }\textbf {\bibinfo {volume} {62}},\ \bibinfo {pages} {214}
  (\bibinfo {year} {1989})}\BibitemShut {NoStop}%
\bibitem [{\citenamefont {Hess}, \citenamefont {Murray},\ and\ \citenamefont
  {Waszczak}(1992)}]{Hess92}%
  \BibitemOpen
  \bibfield  {author} {\bibinfo {author} {\bibfnamefont {H.~F.}\ \bibnamefont
  {Hess}}, \bibinfo {author} {\bibfnamefont {C.~A.}\ \bibnamefont {Murray}}, \
  and\ \bibinfo {author} {\bibfnamefont {J.~V.}\ \bibnamefont {Waszczak}},\
  }\href {\doibase 10.1103/PhysRevLett.69.2138} {\bibfield  {journal} {\bibinfo
   {journal} {Phys. Rev. Lett.}\ }\textbf {\bibinfo {volume} {69}},\ \bibinfo
  {pages} {2138} (\bibinfo {year} {1992})}\BibitemShut {NoStop}%
\bibitem [{\citenamefont {Hess}, \citenamefont {Murray},\ and\ \citenamefont
  {Waszczak}(1994)}]{Hess94}%
  \BibitemOpen
  \bibfield  {author} {\bibinfo {author} {\bibfnamefont {H.~F.}\ \bibnamefont
  {Hess}}, \bibinfo {author} {\bibfnamefont {C.~A.}\ \bibnamefont {Murray}}, \
  and\ \bibinfo {author} {\bibfnamefont {J.~V.}\ \bibnamefont {Waszczak}},\
  }\href {\doibase 10.1103/PhysRevB.50.16528} {\bibfield  {journal} {\bibinfo
  {journal} {Phys. Rev. B}\ }\textbf {\bibinfo {volume} {50}},\ \bibinfo
  {pages} {16528} (\bibinfo {year} {1994})}\BibitemShut {NoStop}%
\bibitem [{\citenamefont {Vlasko-Vlasov}\ \emph {et~al.}(2002)\citenamefont
  {Vlasko-Vlasov}, \citenamefont {Koshelev}, \citenamefont {Welp},
  \citenamefont {Crabtree},\ and\ \citenamefont {Kadowaki}}]{Vlasov02}%
  \BibitemOpen
  \bibfield  {author} {\bibinfo {author} {\bibfnamefont {V.~K.}\ \bibnamefont
  {Vlasko-Vlasov}}, \bibinfo {author} {\bibfnamefont {A.}~\bibnamefont
  {Koshelev}}, \bibinfo {author} {\bibfnamefont {U.}~\bibnamefont {Welp}},
  \bibinfo {author} {\bibfnamefont {G.~W.}\ \bibnamefont {Crabtree}}, \ and\
  \bibinfo {author} {\bibfnamefont {K.}~\bibnamefont {Kadowaki}},\ }\href
  {\doibase 10.1103/PhysRevB.66.014523} {\bibfield  {journal} {\bibinfo
  {journal} {Phys. Rev. B}\ }\textbf {\bibinfo {volume} {66}},\ \bibinfo
  {pages} {014523} (\bibinfo {year} {2002})}\BibitemShut {NoStop}%
\bibitem [{\citenamefont {Yasugaki}\ \emph {et~al.}(2002)\citenamefont
  {Yasugaki}, \citenamefont {Itaka}, \citenamefont {Tokunaga}, \citenamefont
  {Kameda},\ and\ \citenamefont {Tamegai}}]{Tamegai02}%
  \BibitemOpen
  \bibfield  {author} {\bibinfo {author} {\bibfnamefont {M.}~\bibnamefont
  {Yasugaki}}, \bibinfo {author} {\bibfnamefont {K.}~\bibnamefont {Itaka}},
  \bibinfo {author} {\bibfnamefont {M.}~\bibnamefont {Tokunaga}}, \bibinfo
  {author} {\bibfnamefont {N.}~\bibnamefont {Kameda}}, \ and\ \bibinfo {author}
  {\bibfnamefont {T.}~\bibnamefont {Tamegai}},\ }\href {\doibase
  10.1103/PhysRevB.65.212502} {\bibfield  {journal} {\bibinfo  {journal} {Phys.
  Rev. B}\ }\textbf {\bibinfo {volume} {65}},\ \bibinfo {pages} {212502}
  (\bibinfo {year} {2002})}\BibitemShut {NoStop}%
\bibitem [{\citenamefont {Oglesby}\ \emph {et~al.}(1994)\citenamefont
  {Oglesby}, \citenamefont {Bucher}, \citenamefont {Kloc},\ and\ \citenamefont
  {Hohl}}]{Oglesby94}%
  \BibitemOpen
  \bibfield  {author} {\bibinfo {author} {\bibfnamefont {C.~S.}\ \bibnamefont
  {Oglesby}}, \bibinfo {author} {\bibfnamefont {E.}~\bibnamefont {Bucher}},
  \bibinfo {author} {\bibfnamefont {C.}~\bibnamefont {Kloc}}, \ and\ \bibinfo
  {author} {\bibfnamefont {H.}~\bibnamefont {Hohl}},\ }\href {\doibase DOI:
  10.1016/0022-0248(94)91287-4} {\bibfield  {journal} {\bibinfo  {journal}
  {Journal of Crystal Growth}\ }\textbf {\bibinfo {volume} {137}},\ \bibinfo
  {pages} {289 } (\bibinfo {year} {1994})}\BibitemShut {NoStop}%
\bibitem [{\citenamefont {Callaghan}\ \emph {et~al.}(2005)\citenamefont
  {Callaghan}, \citenamefont {Laulajainen}, \citenamefont {Kaiser},\ and\
  \citenamefont {Sonier}}]{Sonier05}%
  \BibitemOpen
  \bibfield  {author} {\bibinfo {author} {\bibfnamefont {F.~D.}\ \bibnamefont
  {Callaghan}}, \bibinfo {author} {\bibfnamefont {M.}~\bibnamefont
  {Laulajainen}}, \bibinfo {author} {\bibfnamefont {C.~V.}\ \bibnamefont
  {Kaiser}}, \ and\ \bibinfo {author} {\bibfnamefont {J.~E.}\ \bibnamefont
  {Sonier}},\ }\href {\doibase 10.1103/PhysRevLett.95.197001} {\bibfield
  {journal} {\bibinfo  {journal} {Phys. Rev. Lett.}\ }\textbf {\bibinfo
  {volume} {95}},\ \bibinfo {pages} {197001} (\bibinfo {year}
  {2005})}\BibitemShut {NoStop}%
\bibitem [{\citenamefont {Campbell}, \citenamefont {Doria},\ and\ \citenamefont
  {Kogan}(1988)}]{Kogan88}%
  \BibitemOpen
  \bibfield  {author} {\bibinfo {author} {\bibfnamefont {L.~J.}\ \bibnamefont
  {Campbell}}, \bibinfo {author} {\bibfnamefont {M.~M.}\ \bibnamefont {Doria}},
  \ and\ \bibinfo {author} {\bibfnamefont {V.~G.}\ \bibnamefont {Kogan}},\
  }\href {\doibase 10.1103/PhysRevB.38.2439} {\bibfield  {journal} {\bibinfo
  {journal} {Phys. Rev. B}\ }\textbf {\bibinfo {volume} {38}},\ \bibinfo
  {pages} {2439} (\bibinfo {year} {1988})}\BibitemShut {NoStop}%
\bibitem [{\citenamefont {Zhang}, \citenamefont {Ting},\ and\ \citenamefont
  {Hu}(2004)}]{Zhang04}%
  \BibitemOpen
  \bibfield  {author} {\bibinfo {author} {\bibfnamefont {D.}~\bibnamefont
  {Zhang}}, \bibinfo {author} {\bibfnamefont {C.~S.}\ \bibnamefont {Ting}}, \
  and\ \bibinfo {author} {\bibfnamefont {C.-R.}\ \bibnamefont {Hu}},\ }\href
  {\doibase 10.1103/PhysRevB.70.172508} {\bibfield  {journal} {\bibinfo
  {journal} {Phys. Rev. B}\ }\textbf {\bibinfo {volume} {70}},\ \bibinfo
  {pages} {172508} (\bibinfo {year} {2004})}\BibitemShut {NoStop}%
\bibitem [{\citenamefont {Fulde}(1969)}]{Fulde69}%
  \BibitemOpen
  \bibfield  {author} {\bibinfo {author} {\bibfnamefont {P.}~\bibnamefont
  {Fulde}},\ }in\ \href@noop {} {\emph {\bibinfo {booktitle} {Tunneling
  Phenomena in Solids}}},\ \bibinfo {editor} {edited by\ \bibinfo {editor}
  {\bibfnamefont {E.}~\bibnamefont {Burstein}}\ and\ \bibinfo {editor}
  {\bibfnamefont {S.}~\bibnamefont {Lundqvist}}}\ (\bibinfo  {publisher}
  {Plenum},\ \bibinfo {address} {New York},\ \bibinfo {year} {1969})\ p.\
  \bibinfo {pages} {427}\BibitemShut {NoStop}%
\bibitem [{\citenamefont {Kohen}\ \emph {et~al.}(2006)\citenamefont {Kohen},
  \citenamefont {Proslier}, \citenamefont {Cren}, \citenamefont {Noat},
  \citenamefont {Sacks}, \citenamefont {Berger},\ and\ \citenamefont
  {Roditchev}}]{Kohen06}%
  \BibitemOpen
  \bibfield  {author} {\bibinfo {author} {\bibfnamefont {A.}~\bibnamefont
  {Kohen}}, \bibinfo {author} {\bibfnamefont {T.}~\bibnamefont {Proslier}},
  \bibinfo {author} {\bibfnamefont {T.}~\bibnamefont {Cren}}, \bibinfo {author}
  {\bibfnamefont {Y.}~\bibnamefont {Noat}}, \bibinfo {author} {\bibfnamefont
  {W.}~\bibnamefont {Sacks}}, \bibinfo {author} {\bibfnamefont
  {H.}~\bibnamefont {Berger}}, \ and\ \bibinfo {author} {\bibfnamefont
  {D.}~\bibnamefont {Roditchev}},\ }\href {\doibase
  10.1103/PhysRevLett.97.027001} {\bibfield  {journal} {\bibinfo  {journal}
  {Phys. Rev. Lett.}\ }\textbf {\bibinfo {volume} {97}},\ \bibinfo {pages}
  {027001} (\bibinfo {year} {2006})}\BibitemShut {NoStop}%
\bibitem [{\citenamefont {Brandt}(1981)}]{Brandt81}%
  \BibitemOpen
  \bibfield  {author} {\bibinfo {author} {\bibfnamefont {E.~H.}\ \bibnamefont
  {Brandt}},\ }\href {\doibase 10.1007/BF00117431} {\bibfield  {journal}
  {\bibinfo  {journal} {Journal of Low Temperature Physics}\ }\textbf {\bibinfo
  {volume} {42}},\ \bibinfo {pages} {557} (\bibinfo {year} {1981})}\BibitemShut
  {NoStop}%
\bibitem [{\citenamefont {Hern\'andez}\ and\ \citenamefont
  {L\'opez}(2008)}]{Hernandez08}%
  \BibitemOpen
  \bibfield  {author} {\bibinfo {author} {\bibfnamefont {A.~D.}\ \bibnamefont
  {Hern\'andez}}\ and\ \bibinfo {author} {\bibfnamefont {A.}~\bibnamefont
  {L\'opez}},\ }\href {\doibase 10.1103/PhysRevB.77.144506} {\bibfield
  {journal} {\bibinfo  {journal} {Phys. Rev. B}\ }\textbf {\bibinfo {volume}
  {77}},\ \bibinfo {pages} {144506} (\bibinfo {year} {2008})}\BibitemShut
  {NoStop}%
\bibitem [{\citenamefont {Boaknin}\ \emph {et~al.}(2003)\citenamefont
  {Boaknin}, \citenamefont {Tanatar}, \citenamefont {Paglione}, \citenamefont
  {Hawthorn}, \citenamefont {Ronning}, \citenamefont {Hill}, \citenamefont
  {Sutherland}, \citenamefont {Taillefer}, \citenamefont {Sonier},
  \citenamefont {Hayden},\ and\ \citenamefont {Brill}}]{Boaknin03}%
  \BibitemOpen
  \bibfield  {author} {\bibinfo {author} {\bibfnamefont {E.}~\bibnamefont
  {Boaknin}}, \bibinfo {author} {\bibfnamefont {M.~A.}\ \bibnamefont
  {Tanatar}}, \bibinfo {author} {\bibfnamefont {J.}~\bibnamefont {Paglione}},
  \bibinfo {author} {\bibfnamefont {D.}~\bibnamefont {Hawthorn}}, \bibinfo
  {author} {\bibfnamefont {F.}~\bibnamefont {Ronning}}, \bibinfo {author}
  {\bibfnamefont {R.~W.}\ \bibnamefont {Hill}}, \bibinfo {author}
  {\bibfnamefont {M.}~\bibnamefont {Sutherland}}, \bibinfo {author}
  {\bibfnamefont {L.}~\bibnamefont {Taillefer}}, \bibinfo {author}
  {\bibfnamefont {J.}~\bibnamefont {Sonier}}, \bibinfo {author} {\bibfnamefont
  {S.~M.}\ \bibnamefont {Hayden}}, \ and\ \bibinfo {author} {\bibfnamefont
  {J.~W.}\ \bibnamefont {Brill}},\ }\href {\doibase
  10.1103/PhysRevLett.90.117003} {\bibfield  {journal} {\bibinfo  {journal}
  {Phys. Rev. Lett.}\ }\textbf {\bibinfo {volume} {90}},\ \bibinfo {pages}
  {117003} (\bibinfo {year} {2003})}\BibitemShut {NoStop}%
\bibitem [{\citenamefont {Huang}\ \emph {et~al.}(2007)\citenamefont {Huang},
  \citenamefont {Lin}, \citenamefont {Chang}, \citenamefont {Sun},
  \citenamefont {Shen}, \citenamefont {Chou}, \citenamefont {Berger},
  \citenamefont {Lee},\ and\ \citenamefont {Yang}}]{Huang07}%
  \BibitemOpen
  \bibfield  {author} {\bibinfo {author} {\bibfnamefont {C.~L.}\ \bibnamefont
  {Huang}}, \bibinfo {author} {\bibfnamefont {J.-Y.}\ \bibnamefont {Lin}},
  \bibinfo {author} {\bibfnamefont {Y.~T.}\ \bibnamefont {Chang}}, \bibinfo
  {author} {\bibfnamefont {C.~P.}\ \bibnamefont {Sun}}, \bibinfo {author}
  {\bibfnamefont {H.~Y.}\ \bibnamefont {Shen}}, \bibinfo {author}
  {\bibfnamefont {C.~C.}\ \bibnamefont {Chou}}, \bibinfo {author}
  {\bibfnamefont {H.}~\bibnamefont {Berger}}, \bibinfo {author} {\bibfnamefont
  {T.~K.}\ \bibnamefont {Lee}}, \ and\ \bibinfo {author} {\bibfnamefont
  {H.~D.}\ \bibnamefont {Yang}},\ }\href {\doibase 10.1103/PhysRevB.76.212504}
  {\bibfield  {journal} {\bibinfo  {journal} {Phys. Rev. B}\ }\textbf {\bibinfo
  {volume} {76}},\ \bibinfo {eid} {212504} (\bibinfo {year}
  {2007})}\BibitemShut {NoStop}%
\bibitem [{\citenamefont {Rodrigo}\ and\ \citenamefont
  {Vieira}(2004)}]{Rodrigo04}%
  \BibitemOpen
  \bibfield  {author} {\bibinfo {author} {\bibfnamefont {J.~G.}\ \bibnamefont
  {Rodrigo}}\ and\ \bibinfo {author} {\bibfnamefont {S.}~\bibnamefont
  {Vieira}},\ }\href {\doibase DOI: 10.1016/j.physc.2003.10.030} {\bibfield
  {journal} {\bibinfo  {journal} {Physica C: Superconductivity}\ }\textbf
  {\bibinfo {volume} {404}},\ \bibinfo {pages} {306 } (\bibinfo {year}
  {2004})}\BibitemShut {NoStop}%
\bibitem [{\citenamefont {Guillamon}\ \emph {et~al.}(2008)\citenamefont
  {Guillamon}, \citenamefont {Suderow}, \citenamefont {Guinea},\ and\
  \citenamefont {Vieira}}]{Guillamon08}%
  \BibitemOpen
  \bibfield  {author} {\bibinfo {author} {\bibfnamefont {I.}~\bibnamefont
  {Guillamon}}, \bibinfo {author} {\bibfnamefont {H.}~\bibnamefont {Suderow}},
  \bibinfo {author} {\bibfnamefont {F.}~\bibnamefont {Guinea}}, \ and\ \bibinfo
  {author} {\bibfnamefont {S.}~\bibnamefont {Vieira}},\ }\href {\doibase
  10.1103/PhysRevB.77.134505} {\bibfield  {journal} {\bibinfo  {journal} {Phys.
  Rev. B}\ }\textbf {\bibinfo {volume} {77}},\ \bibinfo {eid} {134505}
  (\bibinfo {year} {2008})}\BibitemShut {NoStop}%
\end{thebibliography}

%

\end{document}